\documentstyle[12pt]{article}
\newcommand{\eq}{\begin{equation}}
\newcommand{\en}{\end{equation}}
\newcommand{\BEF}{\begin{figure}}
\newcommand{\EF}{\end{figure}}
\newcommand{\bea}{\begin{eqnarray}}
\newcommand{\eea}{\end{eqnarray}}

\newcommand{\s}{\sigma}

\newcommand{\inta}{\int d^{2}x\sqrt{g}\,}
\newcommand{\ts}{\tilde{\sigma}}
\newcommand{\degi}{\delta^{G}}
\newcommand{\dego}{\delta^{G}_{1}(\lambda)}
\newcommand{\diga}{\delta^{G}_{2}(\lambda)}
\newcommand{\del}{\delta^{L}}
\newcommand{\ded}{\delta^{D}}
\newcommand{\lb}{\lambda}
\newcommand{\J}{\J_{\mu ab}}
\newcommand{\T}{\T^{\mu\nu}}
\newcommand{\sdag}{\,^{\dagger}}
\newcommand{\semi}{;\hfil\break}

\newcommand{\epi}{{\epsilon^{\mu\nu} \over \sqrt{g}}}
\newcommand{\unity}{1\kern-.65mm \mbox{\form l}}
\newfont{\form}{cmss10}

\begin{document}
\setlength{\baselineskip}{12pt}
\begin{flushright}
June 1994
\end{flushright}

\vspace{2cm}
\begin{center}
{\Large\bf
$\zeta$-function calculation of the Weyl determinant \\
for two-dimensional non-abelian gauge theories in a curved
background and its W-Z-W terms}

\vspace{1cm}
{\bf Luca Griguolo}\\
{\it International School for Advanced Studies\\
via Beirut 2, 34100 Trieste, Italy\\
I.N.F.N. Sezione di Trieste\\}
\end{center}

\vspace{1cm}
\begin{abstract}
Using a cohomological characterization of the consistent and the
covariant Lorentz and gauge anomalies, derived from the complexification
of the relevant algebras, we study in $d=2$ the definition of the Weyl
determinant for a non-abelian theory with Riemannian background. We
obtain two second order operators that produce, by means of
$\zeta$-function regularization, respectively the consistent and the
covariant Lorentz and gauge anomalies, preserving diffeomorphism
invariance. We compute exactly their functional determinants and the W-Z-W
terms: the ``consistent'' determinant agrees with the non-abelian
generalization of the classical Leutwyler's result, while the
``covariant'' one gives rise to a covariant version of the usual
Wess-Zumino-Witten
action.
\end{abstract}

\newpage
\section*{\normalsize\bf 1.\quad Introduction}
\hspace*{3ex}

%\section{Introduction}
\normalsize
In the first eighties it was realized that a rich algebraic and
geometrical structure subtends the existence of consistent anomalies:
their
characterization as solutions of a cohomological problem \cite{rfb} and its
relation with deep algebraic geometrical theorems, as the index theorem
\cite{rfc,rfu}, were crucial tools in understanding interesting phenomena.
\newline
A complementary approach to the problem of the anomaly relies on the
construction of some representations of the anomaly algebra: with regard
to theories
describing gauge or gravitational interactions of spin one-half fermions,
this is realized as the determinant of a relevant Weyl operator.
\newline
On a compact Riemannian manifold the eigenvalue problem for this
operator is not well defined and the local anomaly is just the
manifestation of this fact: the determinant can be characterized, from a
geometrical point of view, as a line-bundle over the space of the gauge
orbit \cite{rfc}. The non-triviality of this line-bundle, linked to the
non-vanishing of the anomaly, makes impossible to have a global section i.e.
to have gauge invariance for the determinant.

In an analytical approach, where one essentially tries to give a meaning
to an infinite product of eigenvalues, the problem appears from the very
beginning, any possible definition of the
eigenvalues of the Weyl operator
being ambiguos by a phase factor \cite{rfu}. For a gauge theory with
connections on trivial principal-bundle, one can easily by-pass the problem
working with a {\it Dirac} operator, in which the gauge fields are coupled
only to the left or the right component of the spinor.
$\zeta$-function regularization \cite{rfat,rff} provides a well-defined
procedure to define the determinant of operators of this type, reproducing the
perturbative and cohomological results for the consistent local anomaly;
in the gravitational case this trick
does not work, due to the crucial presence of the n-bein fields
that couple to both components.
An analytical definition and an explicit computation of
the chiral determinant in curved space appeared in \cite{rfh,rfi}, using a
completely different approach based on integrating polynomials and a
regularization of the fermion propagator.
\newline
In reference \cite{rfbar} we have developed a systematic way to define the Weyl
determinant for gauge theories, based on its
$\zeta$-function definition in terms of second order operators. The
complexification of the gauge group was the main tool in this approach;
we have shown how the related cohomological problem for the anomaly has
in this case two different solutions, one giving the consistent anomaly
and the other one generating the covariant anomaly. The correct
operators were obtained requiring that the gauge variations of their
functional determinants, in this generalized setting, realize the consistent
solution. In this way we have also clarified the structure of the
covariant anomalies and their relation with functional determinant
techniques.
\newline
In this paper we compute the determinant for a two-dimensional
non-abelian chiral gauge theory in a curved Riemannian background: the
complexification of the gauge groups (color $SU(N)$ and local-frame
$SO(2,R)$) can be avoided, simplifying the analytical calculations, and a
direct comparison of our formalism with previous results \cite{rfi} is
possible. We could also directly test the properties of the covariant
solution and indeed we give the explicit form of the
covariant W-Z-W term \cite{rfd1}, obtained from an
extended $SL(2,C)$ transformation on the functional determinant.
\newline
The paper is organized as follows: to make it self-contained we describe
in Sect.2 the extended cohomological problem in $d=2n$ dimension; in
Sect.3 we turn our attention to $d=2$, writing the anomalies as
derived from the descent equations and explaining
the operatorial approach; in Sect.4 and Sect.5 we find
the correct second order operators linked to the covariant and consistent
anomaly, respectively, and we compute their determinants and the related
W-Z-W
terms.
\section*{\normalsize\bf 2.\quad The extended cohomological problem and
its solution}
The classical action for a chiral spinor coupled to a gauge connection on
a Riemannian manifold $M$ of dimension $2n$ is:
\bea
S_{cl}&=&\int d^{2n}x\sqrt{g}\,\bar{\psi}\,D\,\psi, \label{eq:prima}
\nonumber\\
D&=&e^{\mu}_{a}\,\s_{a}i\,(\partial_{\mu}+iA_{\mu}
+\frac{1}{4}\Omega_{\mu cd}\ts_{c}\s_{d}).\label{eq:seconda}
\eea
We have used the canonical volume form of $M$
\[d^{2n}x \, \sqrt{g},\]
$g_{\mu\nu}$ being a fixed metric. $E_{\mu a}$ are the
$2n$-bein fields (with inverse $e^{\mu}_{a}$) and $\Omega_{\mu cd}$ is the
spin-connection linked to the metric tensor
\begin{eqnarray}
\label{eq:quarta}
g_{\mu\nu}&=&E_{\mu a}\,E_{\nu a}\nonumber\\
\Omega_{\mu ab}&=&e^{\nu}_{a}\,(\partial_{\mu}E_{\nu b}-\Gamma^{\lambda}
_{\mu\nu}E_{\lambda b}).
\end{eqnarray}
We assume the absence of torsion, so $\Gamma^{\lambda}
_{\mu\nu}$ is the usual Levi-Civita connection.
\eq
\label{eq:quinta}
A_{\mu}=A_{\mu}^{a}T_{a}
\en
is a gauge connection belonging to a (trivial) $SU(N)$ principal bundle
over $M$, $T_{a}$ being a basis for the $SU(N)$ Lie algebra:
\begin{eqnarray}
\label{eq:sesta}
[T_{a},T_{b}]&=&i\,f_{abc}T_{c}\nonumber\\
tr(T_{a}T_{b})&=&{1 \over 2}\delta_{ab}
\end{eqnarray}
The Weyl matrices $\s_{a}$ represent the algebra
\eq
\label{eq:sette}
\s_{a}\ts_{b}+\s_{b}\ts_{a}=2\delta_{ab},
\en
$\ts_{a}$ being the other inequivalent representation, existing for
$d=2n$ \cite{rft}.

The vacuum functional of chiral fermions on this backgrounds is formally
defined as
\eq
\label{eq:otto}
\Gamma[A,E]=-\log \det[D].
\en

Immediately we are faced with a basic difficulty in defining the determinant
for the Weyl operator $D$: it maps a chiral spinor on a spinor of
opposite chirality
\begin{equation}
\label{eq:nove}
D : \Gamma (S_{+})_{M} \rightarrow \Gamma (S_{-})_{M}
\end{equation}
where $\Gamma(S_{+})_{M} \, (\Gamma(S_{-})_{M})$ is the space of right (left)
sections of the vector bundle associated by the Dirac representation to the
spin-principal bundle on $M$. $\Gamma (S_{+})_{M}$ and $\Gamma (S_{-})_{M}$ are
different Hilbert spaces and there is no canonical isomorphism
between them: $D$ does not map a Hilbert space into itself and we have
no canonical way to define a meaningful eigenvalue problem, that is
essential for the construction of the determinant. A way out is to try
to define the Weyl determinant by means of an operator with a good
eigenvalue problem but, in general, some of the classical
properties are lost after this substitution.
In particular the covariance properties of $D$ could be destroyed,
breaking the naive invariance of eq.(\ref{eq:otto}) under gauge,
diffeomorphism and local Lorentz transformations.

Disregarding problems of globality (namely assuming that $M$ is
parallelizable), the invariance properties of eq.(\ref{eq:otto})
derived from the symmetry of the classical action eq.(\ref{eq:prima}), are:

$(i)$ local gauge invariance (local $SU(N)$ transformations)
\bea
\label{eq:dieci}
\degi(\lambda)\,A_{\mu}^{a}&=&\partial_{\mu}\lambda^{a}
+i\,[A_{\mu},\lambda]^{a},\nonumber\\
\degi(\lambda)E_{\mu a}&=&0,
\eea
where we have defined $\lb = \lb_{a}T_{a}$;

$(ii)$ local Lorentz invariance (local $SO(2n,R)$ transformations)
\bea
\label{eq:undici}
\del(\tau)A_{\mu}^a&=&0\nonumber\\
\del(\tau)E_{\mu a}&=&\tau_{ab}\,E_{\mu b}
\eea
where we have defined
$\tau_{ab}=-\tau_{ba}$,  $\tau={1\over 4}\tau_{ab}\ts_{a}\s_{b}$;

$(iii)$ diffeomorphism invariance (local $GL(2n,R)$ tranformations)
\bea
\label{eq:diecidici}
\ded(\xi) A_{\mu}^{ a}&=&\partial_{\mu}\xi^{\nu}\,A_{\nu}^{
a}+\partial_{\nu}A_{\mu}^{ a}\,\xi^{\nu},\nonumber\\
\ded(\xi) E_{\mu a}&=&\partial_{\mu}\xi^{\nu}\,E_{\nu a}+\partial_{\nu}E
_{\mu a}\,\xi^{\nu}.
\eea

If we want to describe these symmetries in the more general case of
non-trivial principal bundles and non parallelizable manifolds, we have
to introduce fixed background connections $A^{0}_{\mu}$ and
$\Omega^{0}_{\mu ab}$ in order to get a global description on the base
manifold \cite{rfd1,rfv1}.

The algebra of $\degi$, $\del$ and $\ded$ is:
\bea
\label{eq:dodici}
&[\degi(\lambda_{1}),\degi(\lambda_{2})]&=\degi([\lambda_
{1},\lambda_{2}]),\nonumber\\
&[\ded(\xi_{1}),\ded(\xi_{2})]&=\ded([\xi_{1},\xi_{2}]_{L}),\nonumber\\
&[\del(\tau_{1}),\del(\tau_{2})]&=\del([\tau_{1},\tau_{2}]),\nonumber\\
&[\ded(\xi),\del(\tau)]&=\del(\xi^{\mu}D_{\mu}\tau),
\eea
where
\[ [\xi_1,\xi_2]^{\mu}_{L}=\xi^{\nu}_{1} \partial_{\nu}\xi^{\mu}_{2}
-\xi^{\nu}_{2} \partial_{\nu}\xi^{\mu}_{1}\]
and $D_{\mu}$ is the covariant derivative on $SO(2n,R)$; all the other
commutators vanish.

If we could be able to construct $\Gamma[A,E]$ preserving the classical
properties of the Weyl operator, the invariance of the
action eq.(\ref{eq:prima}) under the just described symmetries
would have been translated
into
\eq
\label{eq:tredici}
\degi(\lb)\Gamma=\del(\tau)\Gamma=\ded(\xi)\Gamma=0.
\en
A meaningful definition of the vacuum functional, in general, will
modify eq.(\ref{eq:tredici}), generating {\underbar anomalies}:
\bea
\label{eq:quattordici}
\degi(\lb)\Gamma[A,E]&=&a^{G}(\lb), \nonumber\\
\del(\tau)\Gamma[A,E]&=&a^{L}(\tau), \nonumber\\
\ded(\xi)\Gamma[A,E]&=&a^{D}(\xi).
\eea

The algebra eq.(\ref{eq:dodici}) gives strong constraints on the
right-hand side of eqs.(\ref{eq:quattordici}), the so-called Wess-Zumino
consistency conditions:
\bea
\degi(\lb_1)a^{G}(\lb_2)-\degi(\lb_2)a^{G}(\lb_1)&=&a^{G}([\lb_1,\lb_2])
,\nonumber\\
\del(\tau_1)a^{L}(\tau_2)-\del(\tau_2)a^{L}(\tau_1)&=&a^{L}([\tau_1,\tau_2])
,\nonumber\\
\ded(\xi_1)a^{D}(\xi_2)-\ded(\xi_2)a^{D}(\xi_1)&=&a^{D}([\xi_1,\xi_2]_{L})
,\nonumber\\
\ded(\xi)a^{L}(\tau)-\del(\tau)a^{D}(\xi)&=&a^{L}(\xi^{\mu}D_{\mu}\tau),
\label{eq:quindici}
\eea
that are equivalent, after having turned the Lie algebra valued
parameters $\lb$, $\tau$ and $\xi$ into anticommuting ghosts and having
defined the action of $\delta^{G,L,D,}$ on them, to a cohomological
problem \cite{rfb}.
The information encoded in eqs.(\ref{eq:quindici}) is enough
to obtain, up to a global coefficient, the general form of $a^{G}$, $a^{L}$
and $a^{D}$, requiring their locality on the fields and their
derivatives. We remark that the cohomological solutions naturally
implement the arbitrariness of the regularization procedure on the
computation of $\Gamma[A,E]$: $a^{G,L,D}$ are obtained up to
coboundaries of the relevant B.R.S.T. operators that correspond to the
addition of suitables local terms on the right-hand side of
eq.(\ref{eq:otto}).

The solutions of eqs.(\ref{eq:quindici}) are called consistent anomalies
and a correct definition of the Weyl determinant must fullfil them,
according eq.(\ref{eq:quattordici}). In the following we assume
$a^{D}(\xi)=0$ because we are interested in a diffeomorphism invariant
definition of $\Gamma[A,E]$; this choice is compatible with
eqs.(\ref{eq:quindici}) (for parallelizable manifolds) and it can be
always achieved by adding to the vacuum functional
a suitable local (but non-polynomial) Wess-Zumino term on the $2n$-bein
fields \cite{rfr}.

If we extend $SU(N)$ to $SL(N,C)$ and $SO(2n,R)$ to $SO(2n,C)$ we can
derive from eqs.(\ref{eq:quindici}) not only the consistent anomalies but
also the covariant anomalies \cite{rfbar,rfm}.
Let us take for example $\degi$: we make the decomposition
\eq
\label{sedici}
\degi(\lb)=\degi_{1}(\lb)+\degi_{2}(\lb)
\en
with $\degi_{1}(\lb)$ generating gauge transformations of $SU(N)$ type
(now seen as subgroup of $SL(N,C)$) and $\degi_{2}(\lb)$ acting on the
ortogonal invariant complement. The algebra of $\degi_{1}$ and
$\degi_{2}$ is:
\bea
&[\degi_1(\lambda_1),\degi_1(\lambda_2)]& =\degi_{1}
([\lambda_1,\lambda_2]),\nonumber\\
&[\degi_2 (\lambda_1),\degi_2(\lambda_2)]& =
-\degi_{1}([\lambda_{1},
\lambda_{2}]),\nonumber\\
&[\degi_{1}(\lambda_{1}),\degi_{2}(\lambda_{2})]& =\degi_{2}
([\lambda_{1},\lambda_{2}]),
\label{eq:diciasette}
\eea
leading to the W-Z conditions:
\bea
\delta^{G}_{1}(\lambda_{1})a^{G}_{1}(\lambda_{2})-\delta^{G}_{1}
(\lambda_{2})a^{G}_{1}
(\lambda_{1})&=&a^{G}_{1}([\lambda_{1},\lambda_{2}]),\nonumber\\
\delta^{G}_{2}(\lambda_{1})a^{G}_{2}(\lambda_{2})-\delta^{G}_{2}
(\lambda_{2})a^{G}_{2}(
\lambda_{1})&=&-a^{G}_{1}([\lambda_{1},\lambda_{2}]),\nonumber\\
\delta^{G}_{1}(\lambda_{1})a^{G}_{2}(\lambda_{2})-\delta^{G}_{2}
(\lambda_{2})a^{G}_{1}(\lambda_{1})&=&a^{G}_{2}([\lambda_{1},\lambda_{2}]).
\label{eq:diciotto}
\eea
It can be shown \cite{rfbar,rfm} that two non-equivalent local solutions of
eqs.(\ref{eq:diciotto}) are characterized by:
\bea
\label{eq:diciannove}
a^{G}_{1}&=&0, \nonumber\\
a^{G}_{2}&\neq&0
\eea
and
\eq
\label{eq:venti}
a^{G}_{2}=\pm i\,a^{G}_{1}.
\en

The functional form of these solutions is given in \cite{rfbar}: they are local
functionals of the connection $A_{\mu}$ on $SL(N,C)$ and of its derivatives.
Projecting $A_{\mu}$ on $SU(N)$, $a^{G}_{2}$, as derived from
eq.(\ref{eq:diciannove}), is the covariant anomaly,
while $a^{G}_{1}$ in eq.(\ref{eq:venti}) is the consistent anomaly.
No differences arise for the $SO(2n,C)$ sector.

The characterization of the consistent anomaly obtained by
eq.(\ref{eq:venti}) will be our guide in the construction of
$\Gamma[A,E]$: we are looking for a definition of $\det[D]$ compatible
with eq.(\ref{eq:venti}). We will also find a functional
realization of eq.(\ref{eq:diciannove}) that we will show to correspond
to the modulus of the Weyl determinant.

\section*{\normalsize\bf 3.\quad
The two dimensional problem: algebraic and functional approach}
We can now apply the general properties discussed in the previous
section to the two-dimensional situation: some simplifications related
to the peculiarity of $d=2$ take place, as we will see.

The first task is to realize the algebras eqs.(\ref{eq:diciasette}) for
$SL(N,C)$ and $SO(2,C)$. The general form of a $SL(N,C)$ Lie valued
gauge connection is \cite{Kd86}:
\eq
\hat{A}_{\mu}={i \over 2}(\delta_{\mu}^{\nu}-i{\epsilon_{\mu}^{\nu}
\over \sqrt{g}})V^{-1}\partial_{\nu}V+
{i \over 2}(\delta_{\mu}^{\nu}+i{\epsilon_{\mu}^{\nu}
\over \sqrt{g}})U^{-1}\partial_{\nu}U
\label{eq:ventuno}
\en
with $U$ and $V$ taking value in the $SL(N,C)$ group. The projection of
$\hat{A}_{\mu}$ over $SU(N)$ is obtained within the restriction
\eq
U^{\dagger}=V^{-1}
\label{eq:ventidue}
\en
that implies $\hat{A}_{\mu}^{\dagger}=\hat{A}_{\mu}$.

We define the action of $\delta_{1}^{G}$ and $\delta_{2}^{G}$ on $U$ and
$V$ as:
\bea
\delta_{1}^{G}(\lambda)U&=&-i U\lb \,\,\,\,\,\,\, \delta_{2}^{G}(\lambda)U=-
U\lb\nonumber\\
\delta_{1}^{G}(\lambda)V&=&-iV\lb \,\,\,\,\,\,\, \delta_{2}^{G}(\lambda)V=V\lb
\label{eq:ventitre}
\eea
that leads to
\bea
\delta_{1}^{G}(\lambda)\hat{A}_{\mu}&=&\hat{D}_{\mu}\lb,\nonumber\\
\delta_{2}^{G}(\lambda)\hat{A}_{\mu}&=&\epsilon_{\mu}^{\nu}\hat{D}_{\nu}\lb
\label{eq:ventiquattro}
\eea
with
\eq
\hat{D}_{\mu}\lb=\partial_{\mu}\lb+i[\lb,\hat{A}_\mu].
\en
One can easily show that $\dego$ and $\diga$ defined by
eqs.(\ref{eq:ventiquattro}) represent the algebra
eq.(\ref{eq:diciasette}). Following the general procedure based on the
descent equation formalism \cite{rfb}, developed in \cite{rfbar,rfd1},
we get the covariant solution eq.({\ref{eq:diciannove}):
\bea
a_{1}^{G}(\lb)&=&0\nonumber\\
a_{2}^{G}(\lb)&=& {1 \over 4\pi}\inta tr \Bigl[ \Bigl({1\over \sqrt{g}}
\epsilon^{\mu\nu}\partial_{\mu}(\hat{A}_\nu+\hat{A}^{\dagger}_{\nu})+
{i\over
\sqrt{g}}\epsilon^{\mu\nu}[\hat{A}_\mu,\hat{A}^{\dagger}_{\nu}]\Bigr)
\lambda\Bigr].
\label{eq:venticinque}
\eea

In the same way the consistent solution eq.(\ref{eq:venti}) is:
\bea
a^{G}_{1}(\lb)&=&ia^{G}_{2}(\lb)={i \over 4\pi}\inta
tr \Bigl[\Bigl({1\over \sqrt{g}}
\epsilon^{\mu\nu}\partial_{\mu}\hat{A}_{\nu}
-i{\cal D}_\mu \hat{A}^{\mu}\Bigr)\lb\Bigr]\nonumber\\
a^{G}_{1}(\lb)&=&-ia^{G}_{2}(\lb)=-{i \over 4\pi}\inta
tr \Bigl[\Bigl({1\over \sqrt{g}}
\epsilon^{\mu\nu}\partial_{\mu}\hat{A}^{\dagger}_{\nu}
+i{\cal D}_\mu \hat{A}^{\dagger\mu}\Bigr)\lb\Bigr].
\label{eq:ventisei}
\eea
where
\[{\cal D}_{\mu}f^{\nu}=\hat{D}_\mu f^{\nu}+\Gamma_{\mu\rho}^{\nu}f^{\rho}.\]
Under $SU(N)$ projection they reduce to the well known expressions for
the covariant and the consistent anomalies. The Bose factor ${1\over 2}$
is a natural bonus of the canonical normalization of the Lie algebra
valued symmetric polynomials we have used to derive the descent
equations.

To obtain a basis for the representation of $SO(2,C)$ we do not need a
complexification of the zwei-bein $E^{\mu}_{a}$: the abelian character
of the group simplifies the approach. We use the representation of the
Weyl algebra:
\bea
\s_{1}&=&\tilde{\s}_{1}=1 , \nonumber\\
\s_{2}&=&-\tilde{\s}_{2}=i\nonumber\\
\eea
and we define
\bea
E_\mu&=&E^{a}_{\mu}\ts_a \nonumber\\
e^\mu&=&e^{\mu}_{a}\s_a \nonumber\\
\Omega_\mu&=&-{1\over 4}\Omega_{\mu a b}\ts_{a}\s_{b}.
\label{eq:ventisette}
\eea
An $SO(2,C)$ matrix $\Lambda$ admits the factorization
\[ \Lambda=R\,\hat{\Lambda} \]
where $R\in SO(2,R)$ and
\eq
\hat{\Lambda}=\left( \begin{array}{cc} \cosh\phi &  -i\sinh\phi \\
                                      i\sinh\phi &  \cosh\phi  \end{array}
                                                              \right).
\en
The action of $\hat{\Lambda}$ is:
\bea
\hat{\Lambda}E_\mu&=&\exp(\phi) E_\mu\nonumber\\
\hat{\Lambda}e^\mu&=&\exp(-\phi) e^{\mu}\nonumber\\
\hat{\Lambda}\Omega_\mu&=&\Omega_{\mu}+{1\over 2}{\epsilon_{\mu}^{\nu}
\over \sqrt{g}}\partial_{\nu}\phi.
\label{eq:ventotto}
\eea
It is clear that the set of fields
\bea
\hat{E}_\mu&=&\exp(\phi) E_\mu^a\nonumber\\
\hat{e}^\mu&=&\exp(-\phi) e^{\mu}_a\nonumber\\
\hat{\Omega}_\mu&=&\Omega_{\mu}+{1\over 2}{\epsilon_{\mu}^{\nu}
\over \sqrt{g}}\partial_{\nu}\phi.
\label{eq:ventinove}
\eea
now carries a representation of $SO(2,C)$ Lie algebra:
\bea
\del_1(\tau)\hat{E}_{\mu}&=&\tau^{ab}\hat{E}_{\mu}^{b}\nonumber\\
\del_2(\tau)\hat{E}_{\mu}&=&i\tau^{ab}\hat{E}_{\mu}^{b}
\label{eq:trenta}
\eea
derived from
\bea
\del_1(\tau)E^a_\mu&=&\tau_{ab}E^b_\mu\nonumber\\
\del_2(\tau)\phi&=&{i\over 2}\tau_{ab}\ts_a\s_b
\label{eq:trentuno}
\eea
being $\del_1(\tau)\phi=0=\del_2(\tau)E^a_\mu$.

Basically we have recognized that
\[Lie\,SO(2,C)\simeq Lie\,(SO(2,R)\otimes R_{+})\]
and we understand $R_{+}$ as the group of the local dilatations of the
orthogonal frame (conformal transformations).
The projection over $SO(2,R)$ is obtained setting $\phi=0$.

Using the descent equation, that in this case is almost trivial due to
the abelian character of the problem, we get:
\bea
a_1^L(\tau)&=&0\nonumber\\
a_2^L(\tau)&=&{i \over 24\pi}\inta R\,\tau
\label{eq:trentadue}
\eea
and
\eq
a_1^L(\tau)=\pm i a_{2}^{L}(\tau)=\pm\inta\Bigl[ {1\over 48\pi}(R\pm 2\Delta_g
\phi)\pm {i \over 12\pi}{\cal D}_\mu\hat{\Omega}^{\mu}\Bigr]\tau
\label{eq:trentatre}
\en
$R$ being the usual scalar curvature.

We remark again the factor ${1\over 2}$ between the covariant and the
consistent solution and the coboundary
\eq
\pm{i \over 12\pi}\inta {\cal D}_{\mu}\hat{\Omega}^\mu \tau
\label{eq:trentaquattro}
\en
that is essential to obtain $a_1=\pm ia_2$. The abelian character of
$SO(2,R)$ makes somewhat questionable the difference between consistent
and covariant Lorentz anomaly in $d=2$: both are proportional to $R$
(apart from coboundary terms) and the definition \cite{rfr} requires only the
factor ${1\over 2}$. Nevertheless in the extended situation they belong
to two different cohomology classes of $\del=\del_1+\del_2$; as one
could check no local term brings eq.(\ref{eq:trentadue}) into
eq.(\ref{eq:trentatre}).

All those algebraic properties possess a precise counterpart once we try
to derive a meaningful functional from the naive definition
eq.(\ref{eq:otto}) of $\Gamma[A,E]$.

Let us turn our attention to the Weyl operator $D$: a very general way
to construct the determinant of a (pseudo) differential operator of
elliptic type is by means of the $\zeta$-function regularization \cite{rfat}.

Under some assumptions we can define:
\bea
\zeta(s;{\cal A})&=&Tr[{\cal A}^{-s}]=\inta
tr\Bigl[K(x,x;s)\Bigr]\nonumber\\
\det[{\cal A}]&=&\exp\Bigl[-{d\over ds}\zeta(s;{\cal A})\Bigr]_{s=0}
\label{eq:trentacinque}
\eea
where $K(x,y;s)$ is the kernel of the complex power of the elliptic
operator ${\cal A}$ \cite{rfs}
\eq
<x|{\cal A}^{-s}|y>=K(x,y;s)
\label{eq:trentasei}
\en
and $Tr$ is understood as an operatorial trace while $tr$ is a matrix
trace.

$K(x,y;s)$ is related to the more usual heat-kernel $H(x,y;t)$
\cite{rfgg} by:
\eq
K(x,y;s)={1 \over \Gamma(s)}\int_0^\infty dt\, t^{s-1}H(x,y;t).
\en
Unfortunately we cannot apply directly the $\zeta$-function machinery to
the Weyl operator: we fail because there is no way to define its complex
power eq.(\ref{eq:trentasei}) for the lack of a ray of minimal growth
\cite{rfs}. This is the analytic counterpart of the bad definition of the
eigenvalue problem, discussed in sect.1.

To overcome this difficulty, we go back to eq.(\ref{eq:otto}) and we
introduce an
isomorphism $D^{*}$
\eq
D^{*}:\Gamma(S_-)_{M}\rightarrow\Gamma(S_+)_M;
\en
then we define
\eq
\det[D]=\det[D^{*}D].
\label{eq:trentasette}
\en
The operator $D^{*}D$ admits a well defined eigenvalue problem, even if the
covariance properties are, {\it a priori}, lost. Our task is to find such a
$D^{*}$ to recover, under $SL(N,C)$ and $SO(2,C)$ variations, the
consistent and the covariant solutions eq.(\ref{eq:diciannove}),
eq.(\ref{eq:venti}). A good candidate for $D^{*}$ is
\eq
D^{*}(r_1,r_2)=i\hat{\s}_{a}\hat{e}^
{\mu}_{a}[\partial_{\mu}+\frac{1}{4}(1-r_{1})\,\hat{\Omega}_{\mu}+
ir_{2}\,{\hat{A\sdag}}_{\mu}]
\label{eq:trentotto}
\en
with $r_1,r_2\in R$. Taking into account our choice of the Weyl matrices
we get for $r_1=0$, $r_2=1$:
\[D^{*}={D\sdag}\]
and
\eq
\det[D^{*}(0,1)D]=\det[{D\sdag}D]=|\det[D]|^{2}
\label{trentanove}
\en
In this case we lose the phase of the determinant on which the
unextended anomaly relies \cite{rfu}.

One can prove that $D^{*}(r_1,r_2)D$ possesses a ray of minimal
growth and the $\zeta$-function technique is perfectly viable. Defining
\eq
\Gamma^{(r_1,r_2)}[\hat{E};\hat{A},{\hat{A\sdag}}]={1 \over
k(r_1,r_2)}\Bigl[{d\over ds}\zeta(s;D^{*}D)\Bigr]_{s=0}
\label{eq:quaranta}
\en
(the choice of the normalization $k(r_1,r_2)$ will be clarified in the
following) and the anomalies
\bea
a_1^{G}(\lambda)&=&\delta_{1}^{G}(\lambda)\Gamma^{(r_1,r_2)}
\,\,\,\,\,\,\,\,\,\,
a_1^{L}(\tau)=\delta_{1}^{L}(\tau)\Gamma^{(r_1,r_2)}\nonumber\\
a_2^{G}(\lambda)&=&\delta_{2}^{G}(\lambda)\Gamma^{(r_1,r_2)}
\,\,\,\,\,\,\,\,\,\,
a_2^{L}(\tau)=\delta_{2}^{L}(\tau)\Gamma^{(r_1,r_2)}
\label{eq:quarantuno}
\eea
we can find the values of $r_1,r_2$ satisfying eqs. (\ref{eq:diciannove})
and (\ref{eq:venti}) with local $a^{(G,L)}_{1,2}$.

To compute the variations of $\Gamma^{(r_1,r_2)}$ we need the
transformation properties of $D^{*}D$ under
$SL(N,C)$ and $SO(2,C)$: using eq.(\ref{eq:ventiquattro}) and
eq.(\ref{eq:trenta}) one obtains:
\bea
\delta^{G}_{1}(\lambda)\Bigl(D^{*}D\Bigr)&=&i\Bigl(D^{*}D\Bigr)\lb-
ir_2\lb \Bigl(D^{*}D\Bigr)-(1-r_2)\Bigl(D^{*} \lb
D\Bigr)+\nonumber\\
&+&i(1-r_2)r_2\tilde{\s}_{a}\hat{e}^{\mu}_{a}[\lb,{\hat{A\sdag}}_{\mu}]D,
\label{eq:quarantadue}
\eea

\bea
\delta^{G}_{2}(\lambda)\Bigl(D^{*}D\Bigr)&=&\Bigl(D^{*}D\Bigr)\lb+
r_2\lb \Bigl(D^{*}D\Bigr)-(1+r_2)\Bigl(D^{*} \lb
D\Bigr)+\nonumber\\
&+&i(1-r_2)r_2\tilde{\s}_{a}\hat{e}^{\mu}_{a}[\lb,{\hat{A\sdag}}_{\mu}]D,
\label{eq:quarantatre}
\eea

\eq
\delta^L_1(\tau)\Bigl(D^{*}D\Bigr)=(1+r_1)\tau\Bigl(D^{*}D\Bigr)
+\Bigl(D^{*}D\Bigr)\tau - r_1\Bigl(D^{*}\tau D\Bigr),
\label{eq:quarantaquattro}
\end{equation}

\eq
\delta^L_2(\tau)\Bigl(D^{*}D\Bigr)=-i(1+r_1)\tau\Bigl(D^{*}D\Bigr)
-i\Bigl(D^{*}D\Bigr)\tau +(r_1-2)\Bigl(D^{*}\tau D\Bigr).
\label{eq:quarantacinque}
\en

The trace properties of the kernel eq.(\ref{eq:trentasei}) \cite{rfs} allow us
to write:
\bea
a^{G}_{1}(\lb)&=&{1\over k}{i\over 4\pi}\inta tr\Bigl[\Bigl(H_1(r_1,r_2)-
\tilde{H}_1(r_1,r_2)\Bigr)\lb\Bigr](1-r_2)+\nonumber\\
&+&{1\over k}{i\over 4\pi}(1-r_2)r_2\inta {d\over ds}\Bigl[s Tr\Bigl\{
\tilde{\s}_{a}\hat{e}^{\mu}_{a}[{\hat{A\sdag}}_{\mu},i\lb]D\Bigl(D^{*}D
\Bigr)^{-s-1}\Bigr\}\Bigr]_{s=0},
\label{eq:quarantasei}
\eea
\bea
a^{G}_{2}(\lb)&=&{1\over k}{i\over 4\pi}\inta tr\Bigl[\Bigl(H_1(r_1,r_2)-
\tilde{H}_1(r_1,r_2)\Bigr)\lb\Bigr](1+r_2)+\nonumber\\
&+&{1\over k}{i\over 4\pi}(1-r_2)r_2\inta {d\over ds}\Bigl[s Tr\Bigl\{
\tilde{\s}_{a}\hat{e}^{\mu}_{a}[{\hat{A\sdag}}_{\mu},\lb]D\Bigl(D^{*}D
\Bigr)^{-s-1}\Bigr\}\Bigr]_{s=0},
\label{eq:quarantasette}
\eea
\eq
a^{L}_{1}(\tau)={1\over k}{1\over 4\pi}\inta tr\Bigl[\Bigl(H_1(r_1,r_2)-
\tilde{H}_1(r_1,r_2)\Bigr)\tau\Bigr]r_1
\label{eq:quarantotto}
\en
\eq
a^{L}_{2}(\tau)={1\over k}{i\over 4\pi}\inta tr\Bigl[\Bigl(H_1(r_1,r_2)-
\tilde{H}_1(r_1,r_2)\Bigr)r_1+4\Bigl(H_1+\tilde{H}_1\Bigr)\Bigr]\tau,
\label{eq:quarantanove}
\en
$H(r_1,r_2)$ and $\tilde{H}(r_1,r_2)$ being the first coefficients of the
expansion of the heat-kernel eq.(\ref{eq:trentacinque}) for $t\rightarrow
0$ \cite{rfgg},  generated respectively by the operators $D^{*}D$ and
$D\,D^{*}$. Generalizing the technique developed in \cite{rfh} to our case (a
deformation of the operator ${D\sdag}D$) we compute the coefficients
$H_1$ and $\tilde{H}_1$:
\bea
H_1(r_1,r_2)&=& \Bigl[-{1 \over 24}-{1\over 8}r_1
\Bigr]\Bigl(R+2\Delta_{g}\phi \Bigr)-{i\over2}
r_1{\cal D}_{\mu}\hat{\Omega}^{\mu}-\nonumber\\
&-&{1\over 2}\Bigl[{\epsilon^{\mu\nu}
\over\sqrt{g}}\partial_{\mu}(\hat{A}_{\nu}+r_2\hat{A}\sdag_{\nu})+
2ir_{2}{\epsilon^{\mu\nu}
\over\sqrt{g}}\hat{A}\sdag_{\mu}\hat{A}_{\nu}-i{\cal D}^{\mu}
(\hat{A}_{\mu}-r_2\hat{A}\sdag_{\mu})\Bigr],
\label{eq:cinquanta}
\eea
\bea
\tilde{H}_1(r_1,r_2)&=& \Bigl[-{1 \over 24}+{1\over 8}r_1
\Bigr]\Bigl(R+2\Delta_{g}\phi \Bigr)+{i\over2}
r_1{\cal D}_{\mu}\hat{\Omega}^{\mu}-\nonumber\\
&-&{1\over 2}\Bigl[-{\epsilon^{\mu\nu}
\over\sqrt{g}}\partial_{\mu}(\hat{A}_{\nu}+r_2\hat{A}\sdag_{\nu})+
2ir_{2}{\epsilon^{\mu\nu}
\over\sqrt{g}}\hat{A}\sdag_{\mu}\hat{A}_{\nu}+i{\cal D}^{\mu}
(\hat{A}_{\mu}-r_2\hat{A}\sdag_{\mu})\Bigr].
\label{eq:cinquantuno}
\eea

We remark that the definition eq.(\ref{eq:quaranta}) automatically
implements the invariance under diffeomorphism: one can easily check
that
\eq
\delta^{D}(\xi)\Bigl(D^{*}D\Bigr)=\Bigl[\xi^{\mu}\partial_{\mu},D^{*}D
\Bigr];
\label{eq:cinquantadue}
\en
the trace properties of the kernel eq.(\ref{eq:trentacinque}) implies
that this type of transformations does not change the $\zeta$-function
determinant.

Let us now look for the covariant and the consistent solution.

\section*{\normalsize\bf 4.\quad The covariant determinant and its W-Z-W
term}
A local expression for $a^{G}_{2}(\lb)$ and $a^{L}_{2}(\tau)$,
compatible with
\[a^{G}_{1}(\lb)=0=a^{L}_{1}(\tau),\]
is obtained by choosing
\bea
r_1&=&0\nonumber\\
r_2&=&1
\label{eq:cinquantatre}
\eea
that corresponds to take $D^{*}=D\sdag$; the normalization $k$ is fixed
to ${1\over 2}$ to recover:
\eq
\Gamma^{(0,1)}\Bigl[\hat{E};\hat{A},\hat{A}\sdag \Bigr]=-\log|\det[D]\,|.
\label{eq:cinquantaquattro}
\en
The explicit form of $H_1$ and $\tilde{H}_1$ leads to the anomalies:
\bea
a^{G}_{2}(\lb)&=&{1\over 4\pi}\inta tr\,\Bigl[\Bigl\{{\epsilon^{\mu\nu}
\over
\sqrt{g}}\partial_\mu(\hat{A}_{\nu}+\hat{A}\sdag_{\nu})+\nonumber\\
&+&i{\epsilon^{\mu\nu}
\over
\sqrt{g}}[\hat{A}_{\mu}\hat{A}\sdag_{\nu}]-i{\cal D}^{\mu}
(\hat{A}_{\mu}-\hat{A}\sdag_{\mu})\Bigr\}\lb\Bigr],
\label{eq:cinquantacinque}
\eea
\eq
a^{L}_{2}(\tau)={i\over 24\pi}\inta\,(R+2\Delta_{g}\phi)\tau,
\label{eq:cinquantasei}
\en
that for $\hat{A}_\mu=\hat{A}\sdag_{\mu}$ and $\phi=0$ become the usual
covariant gauge and Lorentz anomalies:
\bea
a^{G}_{cov}(\lb)&=&{1\over 4\pi}\inta\,tr\Bigl[(\epi
F_{\mu\nu})\lb\Bigr],\cr\label{cinquantasette}
a^{L}_{cov}(\tau)&=&{i\over 24\pi}\inta\,R\tau.
\label{cinquantotto}
\eea
The next step is to compute eq.(\ref{eq:cinquantaquattro}): a first
semplification derives from the fact that $\hat{E}_{\mu}^{a}$ is the
conformal transformed of $E_{\mu}^{a}$, giving us the possibility
of working with the original zwei-bein and, at the end, to recover the
extended result with a local dilatation. A second observation concerns
the diffeomorphism invariance of the determinant itself: we can choose a
convenient coordinate system to develop our calculations.

Locally any two-dimensional Riemannian manifold admits a coordinate
system \cite{rfhi} in which the metric tensor has the form
\eq
g_{\mu\nu}=\exp(4\alpha)\delta_{\mu\nu}
\label{eq:cinquantanove}
\en
leading to the zwei-bein
\eq
e^{\mu}_{a}=\exp(-2\alpha)\delta^{\mu}_{b}\Bigl(\delta_{ab}\cos(2\beta)-
\epsilon_{ab}\sin(2\beta)\Bigr).
\label{eq:sessanta}
\en

Disregarding problems of globality, we assume everywhere the validity of
this parameterization, obtaining a simple expression for the
spin-connection and the scalar curvature:
\bea
\Omega_{\mu}&=&\partial_{\mu}\beta+\epsilon_{\mu\nu}\partial_{\nu}\alpha
\nonumber\\
R&=&-4{1\over\sqrt{g}}\partial_{\mu}\partial_{\mu}\alpha.
\label{eq:sessantuno}
\eea

In these coordinates $D\sdag D$ can be written as
\bea
D\sdag D&=&i\tilde{\s}_{\mu}\exp\Bigl[-3\alpha-i\beta\Bigr]
\Bigl(\partial_{\mu}+i\hat{A}\sdag_{\mu}\Bigr)\exp\Bigl[-2\alpha\Bigr]\cdot
\nonumber\\
&\cdot &i\s_{\nu}\Bigl(\partial_{\nu}+i\hat{A}_{\nu}\Bigr)
\exp\Bigl[\alpha+i\beta\Bigr].
\label{eq:sessantadue}
\eea

It is not difficult to find the infinitesimal variation of $\det[D\sdag
D]$ for the transformations:
\bea
&\alpha&\rightarrow\alpha-\varepsilon\alpha\nonumber\\
&\beta&\rightarrow\beta-\varepsilon\beta
\label{eq:sessantatre}
\eea
with $\varepsilon\rightarrow 0$ and to iterate this change driving
$\alpha$ and $\beta$ to zero:
\bea
\det[D\sdag D]&=&\exp
[\Gamma_1]\det[D\sdag_{2}D_{2}]
\label{eq:sessantaquattro}\nonumber\\
D_{2}&=&i\s_{\mu}\Bigl(\partial_{\mu}+i\hat{A}_{\mu}\Bigr).
\eea

One can easily verify that $\Gamma_1$ actually is only a functional of
$\alpha$ and, eventually, of $\hat{A}_{\mu}$ and $\hat{A}\sdag_{\mu}$:
this fact reflects its invariance under $SO(2,R)$ Lorentz
transformations (independence from the local orthogonal frame, that is
contained in $\beta$, as it is evident from eq.(\ref{eq:sessanta})). In
terms of de-Witt coefficients, $\Gamma_1$ is given \cite{rff} by:
\bea
\Gamma_{1}&=&\int^{1}_{0}dy\,{d\Gamma_1\over dy}(y)\nonumber\\
{d\Gamma_{1}\over dy}(y)&=&{1\over 8\pi}\int
d^{2}x\,\sqrt{g(y)}\Bigl[H_{1}\Bigl(D\sdag(y)D(y)\Bigr)+\tilde{H}_{1}\Bigl(
D(y)D\sdag(y)\Bigr)\Bigr]2\alpha,
\label{eq:sessantacinque}
\eea
where
\[\sqrt{g(y)}=\exp\Bigl[4\alpha(1-y)\Bigr]\]
and $D\sdag(y)D(y)$, $D(y)D\sdag(y)$ are obtained from
eq.(\ref{eq:sessantadue}) with the substitution
\bea
&\alpha&\rightarrow\alpha(1-y),\nonumber\\
&\beta&\rightarrow\beta(1-y).\nonumber\\
\eea

The heat-kernel coefficients are derived from eq.(\ref{eq:cinquanta}) and
eq.(\ref{eq:cinquantuno}):
\eq
\Gamma_{1}[\alpha]={1\over
24\pi}\int\,d^2x\int^{1}_{0}\,dy\,\,(1-y)\,
4(\partial_{\mu}\partial_{\mu}\alpha)\,\alpha,
\label{eq:sessantasei}
\en
that in covariant form is:
\eq
\Gamma_{1}={1\over
192\pi}\inta\,d^{2}z\sqrt{g}\,R(x)\Delta^{-1}_{g}(x,z)R(z),
\label{eq:sessantasette}
\en
$\Delta^{-1}_{g}(x,z)$ being the kernel of the inverse Beltrami-Laplace
operator.

We still have to compute
\[\det\Bigl[D\sdag_{2}D_{2}\Bigr]=\det\Bigl[-\Delta\Bigr]\exp-\Gamma_{2}\Bigl
[\hat{A},\hat{A}\sdag\Bigr],\]
$\det\Bigl[-\Delta\Bigr]$ giving the natural normalization to the free
case (no gauge or gravitational background).

At this point we observe that
\bea
{1\over
2}\s_{\mu}(\delta_{\mu\nu}-i\epsilon_{\mu\nu})&=&\s_{\nu},\nonumber\\
{1\over
2}\tilde{\s}_{\mu}(\delta_{\mu\nu}+i\epsilon_{\mu\nu})&=&\tilde{\s}_{\nu},
\nonumber\\
\eea
while
\[{1\over 2}\s_{\mu}(\delta_{\mu\nu}+i\epsilon_{\mu\nu})=
{1\over 2}\tilde{\s}_{\mu}(\delta_{\mu\nu}-i\epsilon_{\mu\nu})=0;\]
recalling the form of the $\hat{A}_{\mu}$ connection given in
eq.(\ref{eq:ventuno}) we get:
\bea
D\sdag_{2}D_{2}&=&i\tilde{\s}_{\mu}\Bigl[\partial_{\mu}-V\sdag\partial_{\mu}
(V\sdag)^{-1}\Bigr]\,i\s_{\nu}
\Bigl[\partial_{\nu}-V^{-1}\partial_{\nu}V\Bigr]\nonumber\\
&=&V\sdag(i\tilde{\s}_{\mu}\partial_{\mu})(VV\sdag)^{-1}(i\s_{\nu}
\partial_{\nu})V.
\label{eq:sessantotto}
\eea
If we define an interpolating matrix $V(r)$ with the property
\bea
V(0)&=&\unity ,\nonumber\\
V(1)&=&V,
\label{eq:sessantanove}
\eea
we can express, by means of the same decoupling technique used on
eq.(\ref{eq:sessantacinque}), $\Gamma_{2}$ as:
\bea
\Gamma_{2}\Bigl[V,V\sdag \Bigr]&=&\int^{1}_{0}dr\,{d\Gamma_{2} \over dr}
[r;V,V\sdag],\nonumber\\
{d\Gamma_{2}\over dr}[r;V,V\sdag]&=&{1\over 4\pi}\int d^{2}x
\,tr\,\Bigl[\Bigl(H_{1}(r)-\tilde{H}_{1}(r)\Bigr)\Bigl(V^{-1}(r)\partial_{r}
V(r)-\nonumber\\
&-&V\sdag (r)\partial_{r}\Bigl(V\sdag\Bigr)^{-1}(r)\Bigr)\Bigr],
\label{eq:settanta}
\eea
where $H_{1}(r)$ and $\tilde{H}_{1}(r)$ are again derived from
eq.(\ref{eq:cinquanta}) and eq.(\ref{eq:cinquantuno}), taking the pure
gauge part. They give:
\bea
H_{1}(r)-\tilde{H}_{1}(r)&=&-{1\over 2}
\Bigl(\delta_{\mu\nu}+i\epsilon_{\mu\nu}\Bigr)
\partial_{\mu}\Bigl[V\sdag\partial_{\nu}(V\sdag)^{-1}\Bigr](r)+
{1\over 2}
\Bigl(\delta_{\mu\nu}-i\epsilon_{\mu\nu}\Bigr)\partial_{\mu}
\Bigl[V\partial_{\nu}V^{-1}\Bigr](r)-\nonumber\\
&-&{1\over 2}
\Bigl(\delta_{\mu\nu}-i\epsilon_{\mu\nu}\Bigr)
\Bigl[V^{-1}\partial_{\mu}V\,V\sdag\partial_{\nu}(V\sdag)^{-1}+
V\sdag\partial_{\mu}(V\sdag)^{-1}\,V^{-1}\partial_{\nu}V\Bigr].
\label{eq:settantuno}
\eea
It is straightforward to compute the trace on eq.(\ref{eq:settanta})
and to obtain:
\bea
\Gamma_{2}[V,V\sdag]=\hat{\Gamma}[V\,V\sdag]&=&
{1\over 8\pi}\int d^{2}x\,tr\Bigl[\partial_{\mu}\Bigl(V\,V\sdag\Bigr)
\partial_{\mu}\Bigl(V\,V\sdag\Bigr)^{-1}\Bigr]+S_{WZW}\Bigl[V\,V\sdag\Bigr]
\nonumber\\
S_{WZW}[V]&=&{1\over 4\pi}\int^{1}_{0}dr\int d^{2}x\,\epsilon_{\mu\nu}
\Bigl[V^{-1}(\partial_{\mu}V)\,V^{-1}(\partial_{\nu}V)\,V^{-1}(\partial_{r}V)
\Bigr].
\label{eq:settantadue}
\eea
Taking eq.(\ref{eq:settantadue}) in its coordinate invariant form and
performing on eq.(\ref{eq:sessantasette}) a local dilatation, we can
write the extended $D\sdag D$ determinant (normalized to the free case):
\[\det[D\sdag D]=\exp-\Gamma\Bigl[\hat{E};\hat{A},\hat{A}\sdag
\Bigr],\]
\bea
\Gamma\Bigl[\hat{E};\hat{A},\hat{A}\sdag
\Bigr]&=&{1\over
192\pi}\inta\,d^{2}z\sqrt{g}\,R(x)\Delta^{-1}_{g}(x,z)R(z)+\nonumber\\
&+&{1\over 48\pi}\inta\,\Bigl[R\phi+\phi\Delta_{g}\phi\Bigr]+
{1\over 8\pi}\inta\,tr\Bigl[\Bigl(V\,V\sdag\Bigr)^{-1}\Delta_{g}
\Bigl(V\,V\sdag\Bigr)\Bigr]+\nonumber\\
&+& S_{WZW}\Bigl[V\,V\sdag\Bigr].
\label{eq:settantatre}
\eea
We notice that the extra-piece depending on $\phi$ generates a Liouville
action: $\phi$ appears as a dilaton field.

Let us turn our attention to the gauge field part of
eq.(\ref{eq:settantatre}): we recall that $\delta^{G}_{2}$ generates
transformations belonging to $SL(N,C)/SU(N)$. A finite transformation of
this type acts on $V$ and $U$ as:
\bea
&V&\rightarrow V\hat{h},\nonumber\\
&U&\rightarrow U\hat{h}^{-1},\label{eq:settantaquattro}
\eea
where $\hat{h}\in SL(N,C)/SU(N)$. $\Gamma\Bigl[\hat{E};\hat{A},\hat{A}\sdag
\Bigr]$ depends only on the combination $V\,V\sdag$, so the invariance
of eq.(\ref{eq:settantatre}) under $SU(N)$ transformations is evident
while under eq.(\ref{eq:settantaquattro}) we have:
\eq
V\,V\sdag\rightarrow V\hat{h}^{2}V\sdag.
\label{eq:settantacinque}
\en
The well-known Polyakov-Weigmann formula \cite{rfpol}
\eq
\hat{\Gamma}[AB]=\hat{\Gamma}[A]+\hat{\Gamma}[B]+
\inta\,\Bigl(g^{\mu\nu}+i\epi\Bigr)\,tr\Bigl[\Bigl(A^{-1}\partial_{\mu}A
\Bigr)\Bigl(B\partial_{\nu}B\Bigr)\Bigr],
\label{eq:settantasei}
\en
allows us to express the effect of the transformation $\hat{h}$
through a covariant
W-Z-W action \cite{rfd1}:
\[\hat{\Gamma}[V\hat{h}^{2}V\sdag]=\hat{\Gamma}[V\,V\sdag]
+\Gamma^{cov}_{WZW}\Bigl[\hat{h};\hat{A},\hat{A}\sdag\Bigr]\]
\bea
\Gamma^{cov}_{WZW}\Bigl[\hat{h};\hat{A},\hat{A}\sdag\Bigr]&=&2
\hat{\Gamma}\Bigl[\hat{h}\Bigr]+{1\over
4\pi}\inta\,\Bigl(g^{\mu\nu}+i\epi\Bigr)\,tr\Bigl[\hat{A}_{\mu}\hat{A}\sdag_
{\nu}\Bigr]\nonumber\\
&-&{i\over
4\pi}\inta\,\Bigl(g^{\mu\nu}+i\epi\Bigr)\,tr\Bigl[\hat{h}^{-2}
\hat{A}_{\mu}\hat{h}^{2}\hat{A}\sdag_
{\nu}\Bigr]\nonumber\\
&-&{i\over
4\pi}\inta\,\Bigl(g^{\mu\nu}+i\epi\Bigr)\,tr\Bigl[\hat{A}_{\mu}\hat{h}^{2}
\partial_{\nu}\hat{h}^{-2}+\hat{h}^{-2}\partial_{\mu}\hat{h}^{2}\hat{A}\sdag_
{\nu}\Bigr]\nonumber\\
&+&{1\over
4\pi}\inta\,\Bigl(g^{\mu\nu}+i\epi\Bigr)\,tr\Bigl[\hat{h}^{2}
\Bigl(\partial_{\mu}\hat{h}^{-1}\Bigr)\Bigl(\partial_{\nu}\hat{h}^{-1}\Bigr)
\Bigr]\nonumber\\
&+&{1\over
4\pi}\inta\,tr\Bigl[\hat{h}^{2}\Delta_{g}\hat{h}^{-2}-2
\hat{h}\Delta_{g}\hat{h}^{-1}\Bigr].
\label{eq:settantasette}
\eea
$\Gamma^{cov}_{WZW}$ was obtained, in $d=2n$ dimensions, using
cohomological methods in \cite{rfd1}; this is an explicit computation, based
on a functional representation of the relevant extended algebra. The
projection over $SU(N)$ is performed by taking $V\sdag=U^{-1}$. In this
limit $\hat{\Gamma}\Bigl[V\,V\sdag\Bigr]$ coincides with the usual expression
of the logarithm of the Dirac operator, confirming the well known property
\cite{rfu} that
\[|\det[D_{Weyl}]|=\det[D_{Dirac}].\]
Nevertheless by means of an infinitesimal $SL(N,C)/SU(n)$ transformation we can
recover the covariant anomaly as one could directly check in
eq.(\ref{eq:settantasette}) starting from
\[\hat{h}\simeq \unity+\lb.\]
\section*{\normalsize\bf 5.\quad The consistent determinant}
Finally we find the correct second order operator, realizing a mapping
\[\Gamma\Bigl(S_{+}\Bigr)_{M}\rightarrow \Gamma\Bigl(S_{+}\Bigr)_{M}\]
whose $\zeta$-function determinant represents the determinant of the
Weyl operator. In ref.\cite{rfbar}, in the case of pure gravity coupling,
we have
found a whole one-parameter family of second order operators, relaxing
the definition in eq.(\ref{eq:trentasette}), that leads to the
consistent condition eq.(\ref{eq:venti}).

Now we strictly study operators of the form $D^{*}D$, where $D$ is the
Weyl operator: in other words we want to find the correct isomorphism
$D^{*}$ in presence of an additional non-abelian gauge background. We
start from
eqs.(\ref{eq:quarantasei}), (\ref{eq:quarantasette}), (\ref{eq:quarantotto}),
(\ref{eq:quarantanove}), with $k=1$, and we try to fix $r_1$ and $r_2$
leading to:
\bea
a^G_2(\lb)&=&-ia^G_1(\lb),
\label{eq:settantanove}\\
a^L_2(\tau)&=&-ia^L_1(\tau).
\label{eq:ottanta}
\eea
One immediately realizes that $r_2=0$ gives a local expression for
$a^{G}_{1}$ and $a^G_2$ obeying to eq.(\ref{eq:settantanove}). To derive the
correct value of $r_1$ we first switch off the gauge field on
eqs.(\ref{eq:quarantotto}), (\ref{eq:quarantanove}): then using the explicit
form of $H_{1}(r_1,0)$ and $\tilde{H}_{1}(r_1,0)$ we obtain:
\bea
a^{L}_{1}(\tau)&=&{1\over 4\pi}\inta\,\Bigl[-{1\over
4}r_{1}^{2}\Bigl(R+2\Delta_{g}\phi\Bigr)-ir^{2}_{1}{\cal
D}_{\mu}\hat{\Omega}^{\mu}\Bigr]\tau\label{eq:ottantuno},\\
a^{L}_{2}(\tau)&=&{i\over 4\pi}\inta\,\Bigl[-{1\over
4}r_{1}^{2}\Bigl(R+2\Delta_{g}\phi\Bigr)-ir^{2}_{1}{\cal
D}_{\mu}\hat{\Omega}^{\mu}+{1\over
3}\Bigl(R+2\Delta_{g}\phi\Bigr)\Bigr]\tau,
\label{eq:ottantadue}
\eea
forcing the equations:
\bea
{1\over 4}r_{1}^{2}&=&{1\over 3}-{1\over 4}r_{1}^{2}\nonumber\\
r_{1}^{2}&=&-r_{1}^{2}
\label{eq:ottantatre}
\eea
that are clearly inconsistent. But we recall that eq.(\ref{eq:ottanta})
must be understood up to coboundary terms, a fact that corresponds to the
addition of local terms on the logarithm of the determinant
\[ \Gamma^{(r_1,0)}\Bigl[\hat{E};\hat{A},\hat{A}\sdag\Bigr]=
-\log\det\Bigl[D^{*}(r_1,0)D\Bigr] .\]
If we add to the $\zeta$-function calculation of $\Gamma^{(r_1,0)}
\Bigl[\hat{E};\hat{A},\hat{A}\sdag\Bigr]$ the local polynomial
\eq
P[\hat{E}]=-{1\over 12\pi}\inta g^{\mu\nu} \hat{\Omega}_{\mu}
\hat{\Omega}_{\nu}
\label{eq:ottantaquattro}
\en
we have to consider its contribution to $a_1^L(\tau)$ and $a_2^L(\tau)$
that is:
\bea
\delta_{1}^{L}(\tau)P[\hat{E}]&=&-{1\over 6\pi}\inta {\cal
D}_{\mu}\hat{\Omega}^{\mu}\tau,
\label{eq:ottantacinque}\\
\delta_{2}^{L}(\tau)P[\hat{E}]&=&-{i\over 24\pi}\inta
\bigl(R+2\Delta_{g}\phi\Bigr)\tau.
\label{eq:ottantasei}
\eea
These contributions change eq.(\ref{eq:ottantuno}) and
eq.(\ref{eq:ottantadue}), giving a new system that replaces
eq.(\ref{eq:ottantatre}):
\bea
{1\over 2}r_{1}^{2}&=&{1\over 3}-{1\over 6}\nonumber\\
r_{1}^{2}-{2\over 3}&=&-r_{1}^{2}.
\label{eq:ottantasette}
\eea
The two equations are actually the same, leading to the values
\eq
r_1=\pm {\sqrt{3}\over 3},
\label{eq:ottantotto}
\en
that are fully consistent with the general solution of ref.\cite{rfbar}.
The
restoration of the gauge fields $\hat{A}_{\mu}$ and $\hat{A}\sdag_{\mu}$
does not change anything: one can easily verify that their contributions
to $a^L_1(\tau)$ and $a^L_2(\tau)$, for $r_2=0$, automatically satisfy
eq.(\ref{eq:ottanta}), so no new constraints arise from their presence.
To get the consistent gauge and Lorentz anomalies we have only to
compute eq.(\ref{eq:quarantasette}) and eq.(\ref{eq:quarantotto}) with
$r_1={\sqrt{3}\over 3}$ and $r_2=0$:
\bea
a^G_1(\lb)&=&{i\over
4\pi}\inta tr\Bigl[\Bigl(\epi\partial_{\mu}\hat{A}_{\nu}-i{\cal
D}_{\mu}\hat{A}^{\mu}\Bigr)\lb\Bigr],
\label{eq:ottantanove}\\
a^{L}_1(\tau)&=&-{1\over 48\pi}\inta\bigl[R+2\Delta_{g}\phi+4i{\cal
D}_{\mu}\hat{\Omega}^{\mu}\Bigr]\tau.
\label{eq:novanta}
\eea
For $\phi=0$ and $\hat{A}_{\mu}={A}_{\mu}$, we recover the usual form
(up to trivial cocycles), with the correct normalization.

The computation of $\Gamma^{({\sqrt{3}\over 3},0)}
\Bigl[\hat{E};\hat{A},\hat{A}\sdag\Bigr]$ is now straightforward: as in
the previous section we choose the coordinate system
eq.(\ref{eq:sessanta}), where $D^{*}D$
looks like
\bea
D^{*}D&=&\exp\Bigl[\alpha(r_1-3)-i\beta(r_1+1)\Bigr]\Bigl(i\tilde{\s}
\partial_{\mu}\Bigr)\exp\Bigl[-\alpha(r_1+2)+i\beta
r_1\Bigr]\cdot\nonumber\\
&\cdot&D_{2}\exp\Bigl[\alpha+i\beta\Bigr],\nonumber\\
D_{2}&=&i\s_{\nu}\Bigl[\partial_{\nu}-V^{-1}\partial_{\nu}V\Bigr];
\label{eq:novantuno}
\eea
we do not exhibit the explicit calculation, similar to the one described
in the previous section, that leads to
\eq
\Gamma^{({\sqrt{3}\over 3},0)}=\hat{\Gamma}^{({\sqrt{3}\over 3},0)}
-\log\det\Bigl[(i\tilde{\s}_{\mu}\partial_{\mu})D_{2}\Bigr],
\label{eq:novantadue}
\en
\bea
\hat{\Gamma}^{({\sqrt{3}\over 3},0)}&=&{1\over
192\pi}\inta d^{2}z\sqrt{g}\,\Bigl[R(x)\Delta^{-1}_{g}(x,z)R(z)
+iR(x)\Delta^{-1}_{g}(x,z){1\over
\sqrt{g}}\partial_{\mu}\Bigl(\sqrt{g}\Omega^{\mu}(z)\Bigr)
\Bigr]-\nonumber\\
&-&{1\over 24\pi}\inta\Omega_{\mu}\Omega^{\mu},
\label{eq:novantatre}
\eea
where we have taken $\phi=0$ and we have expressed $\hat{\Gamma}
^{({\sqrt{3}\over 3},0)}$ in its diffeo-invariant form. The calculation
of
\[\hat{\Gamma}_{2}=-\log\det\Bigl[(i\tilde{\s}_{\mu}\partial_{\mu})D_{2}\Bigr],
\]
can be directly obtained from eq.(\ref{eq:settantadue}): we notice that
the two determinants are strictly the same if we put $V\sdag=\unity$ in
eq.(\ref{eq:sessantotto}), so that
\eq
\hat{\Gamma}_{2}=\hat{\Gamma}[V];
\label{eq:novantaquattro}
\en
the normalization to the free case is always understood. Summing the two
different contributions
\eq
\Gamma^{({\sqrt{3}\over 3},0)}=\hat{\Gamma}
^{({\sqrt{3}\over 3},0)}+\hat{\Gamma}[V]
\label{eq:novantacinque}
\en
we recognize the non-abelian generalization of the classical
Leutwyler's result \cite{rfi}, with a different choice of the local term in the
gravitational part. The Weyl determinant is obtained as the
$\zeta$-function determinant of the second order operator (we make the
projection on $SU(N)$ and $SL(2,R)$)
\eq
D^{*}D=i\tilde{\s}_{a} e^{\mu}_{a}\Bigl[\partial_{\mu}+(1-{\sqrt{3}\over
3})\Omega_{\mu}\Bigr] i\s_{b}e^{\nu}_{b}\Bigl[\partial_{\nu}+
iA_{\nu}+\Omega_{\nu}\Bigr],
\label{eq:novantasei}
\en
giving an explicit diffeo-invariant result. If we switch off the
zwei-bein field, we recover the well-known form of the two-dimensional
non-abelian gauge determinant \cite{Kd86}.
The usual W-Z-W term is easily obtained by means of a $SU(N)$ gauge
transformation
\[V\rightarrow hV;\]
we notice that a coset trasformation, $\hat{h}\in SL(N,C)/SU(N)$,
leads to the same action for the W-Z-W field.
The characterization we have given of the consistent determinant,
eq.(\ref{eq:venti}), holds also for finite variations:
\bea
&h:&\hat{\Gamma}[V]\rightarrow\hat{\Gamma}[V]+
\Gamma^{con}_{WZW}[h]\nonumber\\
&\hat{h}:&\hat{\Gamma}[V]\rightarrow\hat{\Gamma}[V]+
\Gamma^{con}_{WZW}[\hat{h}].\nonumber\\
\eea

\section*{\normalsize\bf 6.\quad Conclusions}
We have obtained in $d=2$, by characterizing the consistent and the
covariant anomalies as different solutions of an extended cohomological
problem, two second order operators: using $\zeta$-function
regularization they produce, respectively, the covariant and the
consistent Lorentz and gauge anomalies. We have computed their
determinants and the W-Z-W terms linked to the gauge part.

The ``consistent'' determinant is the non-abelian generalization of the
one derived by Leutwyler in \cite{rfi}, as a consequence of a completely
different procedure. If we switch-off the gauge field, the operator in
eq.(\ref{eq:novantasei}) is a particular case of the one-parameter
family of operators, obtained in \cite{rfbar}, that led us to the same
Weyl determinant (up to coboundary terms). At least to our knowledge, in
the case of curved background, it is the first time that the Weyl
determinant has been computed as the $\zeta$-function
determinant of some operator.

The ``covariant'' determinant is essentially the modulus of the Dirac
one: nevertheless working with a $SL(N,C)$ gauge group, this functional
is not invariant under $SL(N,C)/SU(N)$ transformations. The coset action
produces the covariant anomaly in the infinitesimal case, and, for
finite transformations, the covariant W-Z-W term, derived in \cite{rfd1}
by cohomological methods.
\newline
\newline
\newline
I would like to thank Prof. Antonio Bassetto for reading the manuscript
and for many useful comments. I am also grateful to Dott. Pietro Donatis
for having read the manuscript.

\end{document}